\documentclass{preprint}
\usepackage[numbers,comma,sort&compress]{natbib}
\usepackage[T1]{fontenc}
\usepackage{graphicx,verbatim}
\usepackage{xcolor}
\usepackage{booktabs}
\usepackage{colortbl}
\usepackage{subcaption}
\usepackage{marvosym}
\usepackage{tabularx}
\usepackage{makecell}
\usepackage[colorlinks=true, allcolors=blue]{hyperref}
\usepackage{amsmath,amssymb,bm}
\usepackage{pifont}
\title{AMD-Mamba: A Phenotype-Aware Multi-Modal Framework for Robust AMD Prognosis}
\author[1]{Puzhen Wu}
\author[2]{Mingquan Lin}
\author[3]{Qingyu Chen}
\author[4]{Emily Y. Chew}
\author[5]{Zhiyong Lu}
\author[1,*]{Yifan Peng}
\author[1,*]{Hexin Dong}

\begin{document}
%
%
\affil[1]{Department of Population Health Sciences, Weill Cornell Medicine, New York, NY 10022, USA}
\affil[2]{Department of Surgery, University of Minnesota, Minneapolis, MN 55455, USA}
\affil[3]{Department of Biomedical Informatics and Data Science, Yale School of Medicine, New Haven, CT 06510, USA}
\affil[4]{National Eye Institute, National Institutes of Health, Bethesda, MD 20892, USA}
\affil[5]{National Library of Medicine, National Institutes of Health, Bethesda, MD 20892, USA}
\affil[*]{Corresponding author(s). Email(s): \url{yip4002@med.cornell.edu}, \url{hed4006@med.cornell.edu}}

\definecolor{darkgreen}{RGB}{34,139,34}

\newcommand{\yifan}[1]{\textcolor{red}{Yifan: #1}}
\newcommand{\hexin}[1]{\textcolor{blue}{Hexin: #1}}
\newcommand{\puzhen}[1]{\textcolor{darkgreen}{Puzhen: #1}}

\renewcommand{\textbf}[1]{{\fontseries{b}\selectfont #1}}


\maketitle              

\begin{abstract}
Age-related macular degeneration (AMD) is a leading cause of irreversible vision loss, making effective prognosis crucial for timely intervention. 
In this work, we propose AMD-Mamba, a novel multi-modal framework for AMD prognosis, and further develop a new AMD biomarker. This framework integrates color fundus images with genetic variants and socio-demographic variables. At its core, AMD-Mamba introduces an innovative metric learning strategy that leverages AMD severity scale score as prior knowledge. This strategy allows the model to learn richer feature representations by aligning learned features with clinical phenotypes, thereby improving the capability of conventional prognosis methods in capturing disease progression patterns.  
In addition, unlike existing models that use traditional CNN backbones and focus primarily on local information, such as the presence of drusen, AMD-Mamba applies Vision Mamba and simultaneously fuses local and long-range global information, such as vascular changes. Furthermore, we enhance prediction performance through multi-scale fusion, combining image information with clinical variables at different resolutions. 
We evaluate AMD-Mamba on the AREDS dataset, which includes 45,818 color fundus photographs, 52 genetic variants, and 3 socio-demographic variables from 2,741 subjects. Our experimental results demonstrate that our proposed biomarker is one of the most significant biomarkers for the progression of AMD. Notably, combining this biomarker with other existing variables yields promising improvements in detecting high-risk AMD patients at early stages. These findings highlight the potential of our multi-modal framework to facilitate more precise and proactive management of AMD.


\end{abstract}

\section{Introduction}
Age-related macular degeneration (AMD) is a progressive and severe eye disease that primarily affects the macula, the central region of the retina responsible for sharp, detailed vision~\cite{ambati2003age}. The diagnosis of AMD is based mainly on color fundus imaging, and the disease can be generally classified into early, intermediate, and late stages~\cite{age2001age}. In its late stages, AMD can lead to significant central vision loss or even legal blindness, profoundly impacting patients' quality of life~\cite{ferris2013clinical}. Consequently, early detection, prevention, and appropriate management strategies are crucial to slowing AMD progression and preserving vision.

\begin{figure}[tb]
    \centering
    \begin{subfigure}[t]{0.49\textwidth}
    \centering
    \includegraphics[width=0.75\linewidth]{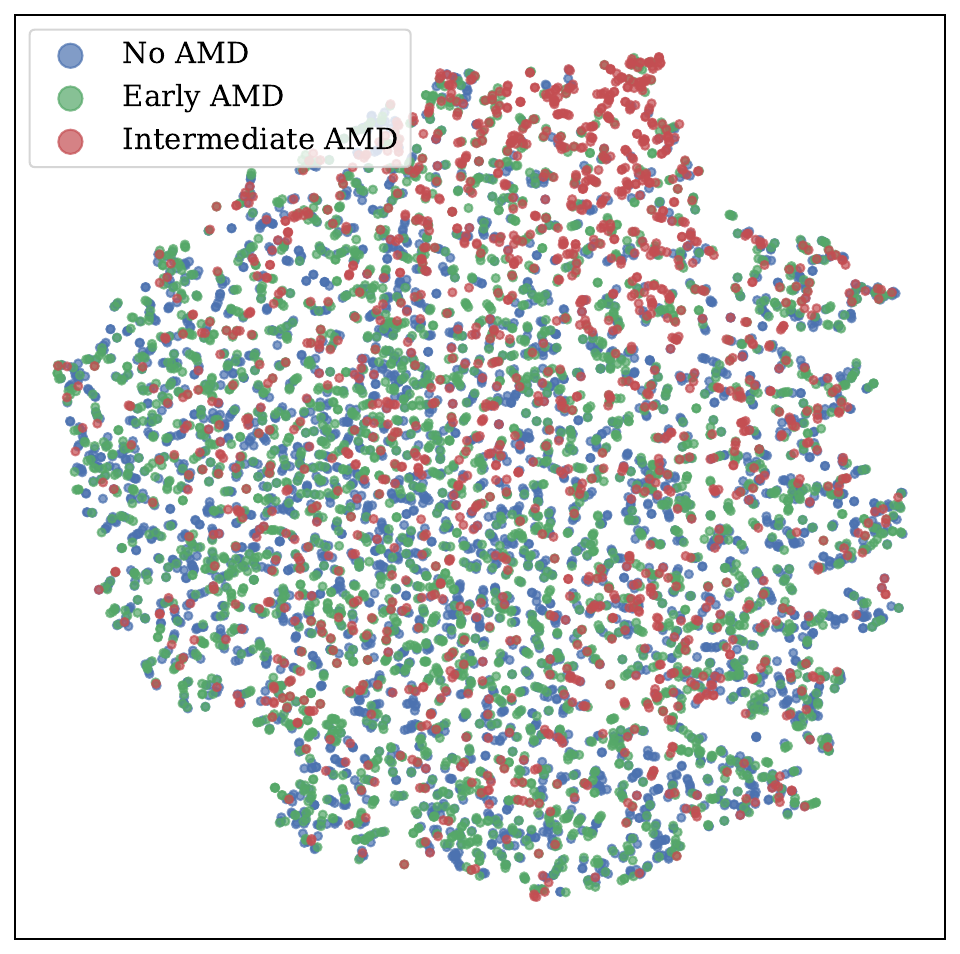}
    \caption{}
    \end{subfigure}
    \begin{subfigure}[t]{0.49\textwidth}
    \centering
    \includegraphics[width=0.75\linewidth]{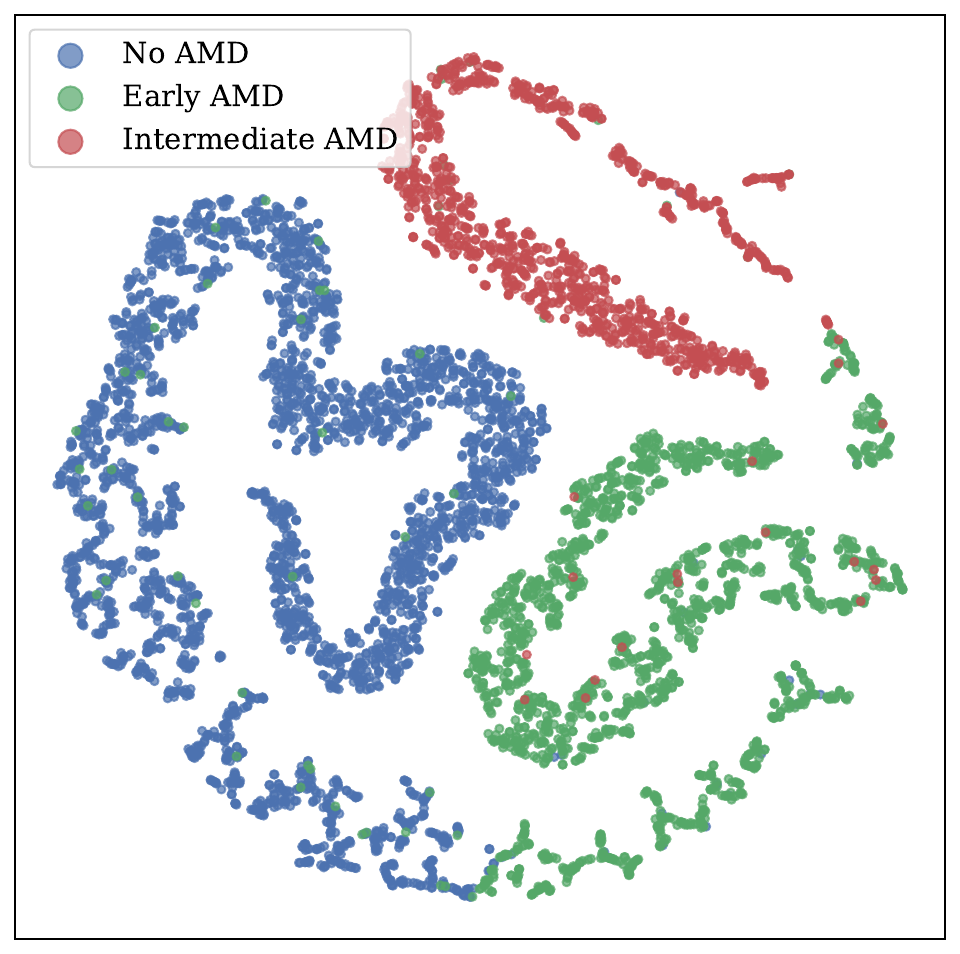}
    \caption{}
    \end{subfigure}
    \caption{T-SNE visualization of learned features comparing \textbf{(a) the previous
    AMD prognosis method}~\cite{yan2020deep} and \textbf{(b) AMD-Mamba}. By incorporating AMD severity score as a key prior, AMD-Mamba results in clusters with clearer separations.}
    \label{fig:fig1}
\end{figure}

In recent years, deep learning models have excelled in classifying AMD categories~\cite{peng2019deepseenet,burlina2017automated,lee2017deep,ting2017development}. However, it is important to recognize that, predicting AMD progression risk is more crucial than merely determining its current stage, as it better guides clinical interventions and treatment planning. Researchers have introduced a variety of prognosis models, including two-stage Cox-based frameworks~\cite{peng2020predicting}, end-to-end k-year survival model~\cite{yan2020deep,babenko2019predicting}, interpretable prognosis model \cite{gervelmeyer2024interpretable}, and longitudinal AMD prognosis model \cite{holste2024harnessing}. Despite these advancements, most methods ignore the AMD phenotype (i.e., step-wise AMD severity scale scores), which are highly related to AMD progression~\cite{ferris2013clinical}. For example, Peng et al.~\cite{peng2020predicting} employed a binary classification model as a pretrain model to classify late AMD, but neglected the transitions between early and intermediate stages. Yan et al.~\cite{yan2020deep} directly used the classification results as the input for survival analysis, disregarding potentially informative image-level texture features. In contrast, we introduce the AMD severity score as a key prior to our survival prognosis model. The proposed method not only reduces dependence on large amounts of labeled data, which is especially relevant given the often limited availability of labeled prognostic datasets, but also enables the model's ability to learn robust texture features that more effectively capture AMD progression (Fig.~\ref{fig:fig1}).

Additionally, most existing AMD prognosis methods rely on CNN-based structures as image encoders~\cite{peng2020predicting,yan2020deep,babenko2019predicting,gervelmeyer2024interpretable,holste2024harnessing}. While CNNs are effective at capturing local features like the presence of drusen, they may struggle with incorporating broader contextual information like vascular changes, which also plays a pivotal role in AMD progression~\cite{coscas2019optical}. Recently, self-attention-based architectures (e.g., ViT~\cite{dosovitskiy2021an}, U-Mamba~\cite{U-Mamba}, and V-Mamba~\cite{liu2024vmamba-l}) have demonstrated substantial success across various vision tasks. Inspired by these architectures, we propose a novel \textbf{AMD-Mamba} architecture that simultaneously addresses local and long-range information. By integrating spatial and channel attention mechanisms, AMD-Mamba adaptively emphasizes crucial local details. In addition, genetic and socio-demographic variables are recognized as key contributors to AMD progression~\cite{fritsche2016large}. Consequently, AMD-Mamba integrates these variables alongside multi-scale image features. Thus, it not only provides a more comprehensive representation of disease risk but also helps guide the network to focus on subtle indicators, such as minor microvascular changes or small-scale drusen growth, which might otherwise be overlooked, ultimately leading to more robust and wide-ranging prognostic predictions.

In this study, our contributions are as follows: 1) \textbf{Incorporation of AMD Phenotype}: We incorporate AMD severity score as a critical prior in our prognostic model. This approach reduces the reliance on extensive labeled data and allows the model to learn more robust features. 2) \textbf{Development of AMD-Mamba Architecture}: It captures local and global information and integrates multi-scale image features with genetic and socio-demographic variables to comprehensively understand AMD progression. 3) \textbf{Development of a New Multi-modal AMD Biomarker}: Leveraging the model's predicted risk, we further develop a new AMD biomarker that remains statistically significant in the multivariate analysis even after adjusting with established clinical predictors\cite{ferris2013clinical}. This biomarker holds promise for enhancing risk stratification and treatment planning for AMD patients. 4) \textbf{Multicenter Verification}: We verify the effectiveness of our approach through 5-fold cross-validation and statistical analyses on the public, multi-center Age-Related Eye Disease Study (AREDS) dataset.

\section{Method}

\subsection{Proposed Architecture}

\begin{figure}[tb]
    \centering
    \includegraphics[width=0.85\textwidth]{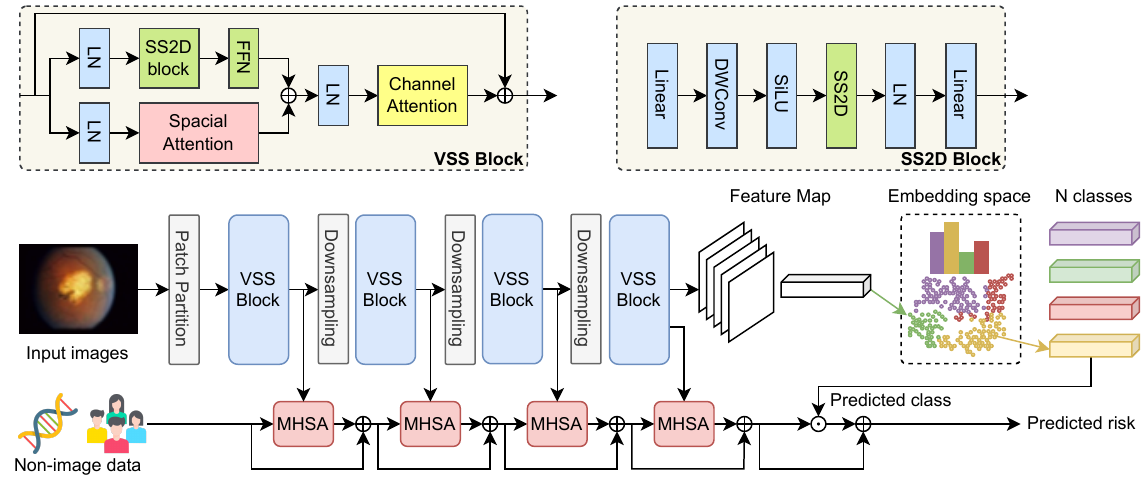}
    \caption{An overview of AMD-Mamba. Stage 1 learns discriminative visual features through classification, guided by AMD severity scores. Stage 2 fuses the frozen backbone's multi-scale outputs with genetic and socio-demographic data via MHSA and predicts progression risk using a survival head.}
    \label{fig:framework}
\end{figure}

Our vision backbone builds upon the V-Mamba~\cite{liu2024vmamba-l}. As shown in Fig.~\ref{fig:framework}, the input image is first processed through a 
patch embedding layer, resulting in high-dimensional token representations. These tokens progress
through a stack of Visual State Space (VSS) blocks interleaved
with downsampling operations. Unlike the blocks in V-Mamba \cite{liu2024vmamba-l}, our proposed block employs a two-branch design to capture both local details and global contextual cues. Specifically, each VSS block takes a feature tensor 
$\mathbf{X}\in\mathbb{R}^{H\times W\times C}$.
We then feed $\mathbf{X}$ into two branches:
the left branch applies LayerNorm followed by a 2D-selective-scan module (SS2D) \cite{liu2024vmamba-l}
and a feed-forward network (FFN), 
while the right branch applies LayerNorm (LN) followed by spatial attention (SA) \cite{woo2018cbam}.
After summing the outputs of these two branches, 
we apply channel attention (CA) \cite{hu2018squeeze} after LayerNorm, 
and finally add the original $X$ as a skip connection:
\begin{equation}
\mathbf{X}_\text{out}
= \mathbf{X} + \text{CA}\Bigl(\text{LN}\bigl(\text{FFN}(\text{SS2D}(\text{LN}(\mathbf{X}))) 
+ \text{SA}(\text{LN}(\mathbf{X}))\bigr)\Bigr)
\end{equation}

As the network progressively reduces spatial resolution and expands channel
dimensionality across multiple VSS blocks and downsampling layers, it yields a 
sequence of 4 multi-scale feature maps \(\{\mathbf{f}_1,\ldots,\mathbf{{f}_4}\}\)
that capture increasingly abstract representations, with \(\mathbf{f}_4\) being the final, 
lowest-resolution feature map. These feature maps serve as key inputs for
the subsequent survival prognosis step, where they are fused with gene-demographic 
information via a multi-head self-attention (MHSA) module \cite{vaswani2017attention}. The fused features are then passed to a survival head, allowing AMD progression prediction.

\subsection{Training strategy}

We apply a two-stage approach. 
Stage 1 learns discriminative visual features through classification, guided by AMD severity scores. Stage 2 fuses the frozen backbone's multi-scale outputs with genetic and socio-demographic data via MHSA and predicts progression risk using a survival head.


\textbf{Stage 1: Metric-driven Classification Pretraining.}
In this stage, our goal is to obtain high‐quality visual features from fundus images through a supervised classification task. We achieve this by using a set of embeddings that enable a metric-based decision rule. Let
$\mathbf{I}\in\mathbb{R}^{H\times W\times3}$ be an input image, and $f(\mathbf{I};\theta_f)$ the
vision backbone producing a latent feature vector $\mathbf{f}_4\in\mathbb{R}^d$. We maintain a
learnable matrix $\mathbf{g}\in\mathbb{R}^{C\times d}$, where $C$ is the number of AMD phenotype categories,  with each row $\mathbf{g}_i$ serving as the
prototype for class $i$. The classification logits $y_i$ are then computed using cosine similarity:
\begin{equation}
    y_i = \cos(\mathbf{f}_4, \mathbf{g}_i)
    = \frac{{\mathbf{f}_4}^\top \mathbf{g}_i}{\|\mathbf{f}_4\| \|\mathbf{g}_i\|},
    \quad i \in \{1,\ldots,C\}
\end{equation}
For a training sample labeled as $y\in\{1,\dots,C\}$, we optimize the network using the 
\emph{cross-entropy loss} based on cosine similarity:
\begin{equation}
    \mathcal{L}_{\text{CE}}(\mathbf{X}, y; \theta_f, \mathbf{g})
    = -\log\left(\frac{\exp\bigl(\cos(\mathbf{f}_4,\mathbf{g}_y)\bigr)}
    {\sum_{j=1}^{C}\exp\bigl(\cos(\mathbf{f}_4,\mathbf{g}_j)\bigr)}\right)
    \label{eq:ce_loss}
\end{equation}
By optimizing $\mathcal{L}_{\text{CE}}$, $\theta_f$ are adjusted so that 
$\mathbf{f_4}$ is closely aligned (in angular distance) with its correct class prototype $\mathbf{g}_y$.
Simultaneously, this process ensures that $\mathbf{g}_y$ effectively represents the cluster of 
training samples belonging to class $y$. Upon the completion of Stage~1, we obtain a pretrained backbone $f(\cdot;\theta_f)$ and a set of 
learned class novels for $C$ AMD phenotype categories \(\{\mathbf{g}_1,\dots,\mathbf{g}_C\}\), both of which are leveraged in Stage~2 
for further survival analysis.

\textbf{Stage 2: Multi-modal Survival Prediction.} \label{sec:stage2}
Here, we freeze the parameters of this 
backbone to preserve its discriminative capacity. Each feature map \(\{\mathbf{f}_1,\ldots,\mathbf{f}_4\}\) is pooled into 
\(\bar{\mathbf{f}}_i \in \mathbb{R}^d\). Meanwhile, we concatenate the genetic and demographic 
vectors into \(\mathbf{e}\in\mathbb{R}^{d_e}\) and project it through a learnable linear 
projection $\mathbf{W}_q \in \mathbb{R}^{d\times d_e}$ 
that maps $\mathbf{e}$ into an initial query embedding $d$-dim query 
$\mathbf{q_1} = \mathbf{W}_q\,\mathbf{e}$. In an MHSA, each \(\bar{\mathbf{f}}_i\) serves as a key-value. Concretely, for each scale \(i\) in ascending order, we define
\(\mathbf{k}_i = \mathbf{W}_k\bar{\mathbf{f}}_i\) and 
\(\mathbf{v}_i = \mathbf{W}_v\bar{\mathbf{f}}_i\), where $\mathbf{W}_k,\mathbf{W}_v \in \mathbb{R}^{d\times d}$ are two learnable linear mappings that project 
$\bar{\mathbf{f}}_i$ into key and value vectors. The fused 
embedding \(\mathbf{q_i}\) is then iteratively updated by:
\begin{equation}
\mathbf{q}_{i+1}
\leftarrow
\mathbf{q}_i
+
\mathrm{MHSA}\bigl(\mathbf{q}_i,\mathbf{k}_i,\mathbf{v}_i\bigr)
\end{equation}
Then, we pass the result through a feed-forward block with skip connections for additional refinement. Once all four scales are processed, the final embedding \(\mathbf{q_4}\) captures multi-resolution cues from the image, genetic, and demographic information. To incorporate the classification output from Stage 1 into our survival analysis, we retain the class-embedding matrix $\mathbf{g}$. This serves as a phenotypic prior that allows our Stage 2 model to emphasize features aligned with the most likely AMD category. Given the lowest-resolution feature map 
\(\mathbf{f}_4\), we compute 
\(\hat{s} = \arg\max_{c}{\cos(\mathbf{f}_4,\mathbf{g}_c)}\) 
to determine the most likely class prototype \(\mathbf{g}_{\hat{s}}\), where \(c \in \{1, \dots, C\}\). We then combine 
\(\mathbf{g}_{\hat{s}}\) with the fused embedding \(\mathbf{q_4}\) 
via an elementwise product, followed by a skip connection:
\begin{equation}
\mathbf{u}^* 
=
\mathbf{q_4} 
+
\bigl(\mathbf{q_4}\odot\mathbf{g}_{\hat{s}}\bigr)
\end{equation}
Thus,  the original fused representation is preserved  while selectively emphasizing features 
aligned with the predicted class. Finally, \(\mathbf{u}^*\) is passed to a shallow MLP 
to predict the log-risk \(\beta\). The parameters of this survival head are optimized 
under a negative Cox partial log-likelihood \cite{cox1972regression}.
\begin{equation}
\label{eq:cox-loss}
\mathcal{L}(\boldsymbol{\beta}) =- \sum_{i:\delta_i=1}\Bigl(\beta_i-\log\sum_{j\in R(t_i)} \exp{\beta_j}\Bigr)
\end{equation}
\(\delta_i\) indicates 
whether subject \(i\) is uncensored, and \(R(t_i)\) is the risk set at time \(t_i\). 

\section{Experiments and Results}

\textbf{Datasets.} We evaluate our method on the publicly available Age-Related Eye Disease Study (AREDS) dataset(\url{https://www.ncbi.nlm.nih.gov/projects/gap/cgi-bin/study.cgi?study\_id=phs000001.v3.p1})~\cite{ferris2013clinical}.  Due to the publicly available nature of AREDS, the requirement for obtaining written informed consent from all subjects was waived by the IRB. AREDS contains 45,818 color fundus images from 2,741 subjects, along with 3 socio-demographic variables (age, sex, and smoking status) and 52 genetic variants derived from~\cite{yan2020deep} (Table~\ref{tab:data}). Each image is assigned an AMD severity score between 1 and 12, with scores of 10 or higher indicating late AMD. We group these scores into four classes: no (score=1), early (scores 2–5), intermediate (scores 6–9), and late (scores 10–12) AMD. In Stage~1, we use all 45,818 color fundus images for classification pretraining. In Stage~2, we focus on the 4,977 images from the base visit of eyes without late AMD (score < 10) for survival analysis.
\begin{table}[tb]
\caption{Characteristics of AREDS.}
\label{tab:data}
\centering
\begin{tabular}{l@{~~~}ll}
\toprule
\multicolumn{3}{l}{Participants characteristics:}  \\
& Number of participants & 2,741\\
& Age, mean (SD) & 73.9 (4.9)\\
& Sex (F/M) & 1,545/1,196\\
& Smoking status (never/former/current) & 1,287/1,284/170\\
\midrule
\multicolumn{3}{l}{Color fundus images:}\\
& Images for pretraining (Stage 1) & 45,818\\
& Images from the base visit (Stage 2) & 4,977\\
& AMD severity scale score from the base visit&       \\
& \;\;\;(no/early/intermediate) & 2,189/1,973/815\\
& Progression to late AMD (all years):  & 584\\
\bottomrule
\end{tabular}
\end{table}

\textbf{Experimental Details.} All experiments run on an NVIDIA RTX A6000 GPU with a 5-fold split (by patient ID) of the AREDS dataset. Images are resized to $224\times224$ pixels and then augmented via random $\pm10^\circ$ rotation, horizontal flipping ($p=0.5$), and normalized using ImageNet statistics. 
In Stage 1, we use the Adam optimizer (learning rate $10^{-4}$, batch size 96) for 50 epochs, and in Stage 2, the same optimizer settings are employed (learning rate $10^{-4}$, batch size 512) for 100 epochs. We select the best model based on the validation C-index.

\textbf{Ablation Study.} An ablation study is conducted to assess the impact of various design choices (Table~\ref{tab:ablation}).  First, adding clinical variables to the original Mamba architecture (M1) improved the C‐index from 0.8634 to 0.8713, confirming the benefit of those variables. Integrating multi‐scale attention (M3) further boosts the C‐index to 0.8781, highlighting the importance of capturing both local and global features. Extending M3 with a ``hard label'' strategy (M4), where the class with the highest predicted probability from Stage 1 is selected and multiplied elementwise with \(\mathbf{q_4}\), raises the C-index to 0.8873. Alternatively, using a ``soft label'' approach (M5), which weights each class by its probability for elementwise multiplication with \(\mathbf{q_4}\), resulted in a slightly lower C-index of 0.8810. Finally, replacing the Mamba backbone with DenseNet in the best-performing setting (M6) achieved a C-index of 0.8729, underscoring Mamba's advantage. When comparing M7 and M8, using either the original 12 phenotypic categories (12c) or a simple binary label separating late AMD from no AMD (2c) resulted in lower performance than our four-category approach, indicating that a balanced division of AMD stages is crucial for accurately capturing progression.

These findings demonstrate that each proposed design component -- backbone choice, multi‐scale feature extraction, fusion strategy, and label guidance -- significantly improves prognostic accuracy.

\begin{table}[tb]
\centering
\caption{Ablation study demonstrating the impact of different backbones, multi‐scale attention, tabular data fusion, and label guidance strategies.}
\label{tab:ablation}
\begin{tabular}{clllc}
\toprule
Models & Backbone & Clinical variable fusion         & Label guidance        & C-index \\ \midrule
1  & Mamba    & -                     & -                   &0.8634 $\pm$ 0.0126         \\ 
2  & Mamba    & Concat Fusion         & -                   &0.8713 $\pm$ 0.0110         \\ 
3  & Mamba    & multi-scale attention & -                   &0.8781 $\pm$ 0.0158         \\ 
4  & Mamba    & multi-scale attention & hard label          &\textbf{0.8873} $\pm$ 0.0093         \\ 
5  & Mamba    & multi-scale attention & soft label          &0.8810 $\pm$ 0.0080        \\ 
6  & DenseNet & multi-scale attention & hard label     &0.8729 $\pm$ 0.0097         \\ 
7  & Mamba    & multi-scale attention & 12c hard label &0.8795 $\pm$ 0.0092          \\ 
8  & Mamba    & multi-scale attention & 2c hard label  & 0.8807 $\pm$ 0.0104         \\ \bottomrule
\end{tabular}
\end{table}

\textbf{Comparisons with SOTA.} Table~\ref{tab:comparison} compares our proposed approach against several previous methods using a 5-fold cross-validation setting. Unlike some existing works that exclusively rely on image data, our approach integrates relevant tabular information, such as genetic variants and socio-demographic variables. This integration achieves a superior C-index of 0.8873 and a 5-year AUC of 0.8942, surpassing both image-only and other multi-modal baselines. These results underscore the benefits of incorporating multi-modal data for a more accurate AMD prognosis.

\begin{table}[tbp]
\newcommand{\cmark}{\ding{51}}%
\newcommand{\xmark}{\ding{55}}%
\centering
\caption{Results of different methods under 5‐fold cross‐validation.}
\label{tab:comparison}
\begin{tabular}{lccccc}
\toprule
               & Image & Genetic & Socio-demo. & {C-Index} & {5-years AUC} \\ \midrule
Babenko et al.~\cite{babenko2019predicting} &  \cmark & \xmark & \xmark &      -- &  0.8399 $\pm$ 0.0287   \\ 
Yan et al.~\cite{yan2020deep}     &  \cmark & \xmark & \xmark & -- &  0.8401 $\pm$ 0.0375   \\   
BagNet~\cite{gervelmeyer2024interpretable}   &  \cmark & \xmark & \xmark & 0.8241 $\pm$ 0.0151 & 0.8362 $\pm$ 0.0044 \\ 
Ours           & \cmark & \xmark & \xmark &\textbf{0.8634} $\pm$ 0.0126                 & \textbf{0.8717} $\pm$ 0.0135            \\ \midrule
Peng et al.~\cite{peng2020predicting}    &  \cmark   &  \cmark    &  \cmark  & 0.8337 $\pm$ 0.0149   &  0.8419 $\pm$ 0.0106  \\  
Yan et al.~\cite{yan2020deep}    & \cmark    & \cmark  & \xmark  & --  & 0.8449 $\pm$ 0.0164   \\   
Ours     &   \cmark     &   \cmark     & \cmark &   \textbf{0.8873} $\pm$ 0.0093     &  \textbf{0.8942} $\pm$ 0.0107           \\ \bottomrule
\end{tabular}
\end{table}
\begin{table}[!htb]
\caption{Multivariate Cox regression analysis. Variables with p-values of 0.05 or lower are shown. HR: hazard ratio. SUBFF2: Subretinal fibrosis field 2 (yes/no). RPEDWI: RPE Depigmentation area w/i grid (0-8).}
\centering
\begin{tabular}{lrrr}
\toprule
Variables  & HR  & (95\% CI) & p-value \\ \midrule
Ours  & \textbf{2.77} & (2.00~3.82) & <0.005 \\ 
AMD score & 1.59 & (1.45~1.73) & <0.005   \\ 
\midrule
\multicolumn{4}{l}{\cellcolor{gray!20}Phenotype}     \\ 
SUBFF2 & 0.21 & (0.09~0.48) & <0.005  \\
RPEDWI & 1.06 & (1.01~1.12) & 0.0288 \\ 
\midrule
\multicolumn{4}{l}{\cellcolor{gray!20}Socio-demographic}  \\ 
Age & 1.30 & (1.03~1.65) & 0.0283  \\
Smoking status & 1.17 & (1.02~1.34) & 0.0278 \\ 
\midrule
\multicolumn{4}{l}{\cellcolor{gray!20}Genetic variants}\\
rs10922109\_A & 0.81 & (0.67~0.99) & 0.0390     \\
rs121913059\_T & 2.10 & (1.07~4.12) & 0.0320\\
rs140647181\_C & 1.79 & (1.19~2.69) & 0.0051\\
rs114092250\_A & 0.47 & (0.26~0.87) & 0.0167  \\
rs116503776\_A & 0.73 & (0.58~0.92) & 0.0069\\
rs3750846\_C & 1.30 & (1.15~1.47) & <0.005\\
rs9564692\_T & 0.87 & (0.76~1.00) & 0.0472      \\
rs61985136\_C & 0.87 & (0.77~0.99) & 0.0404 \\    
           \bottomrule
\end{tabular}

\label{tab:multicox}
\end{table}

\begin{figure}[tb]
    \centering
    \includegraphics[width=\textwidth]{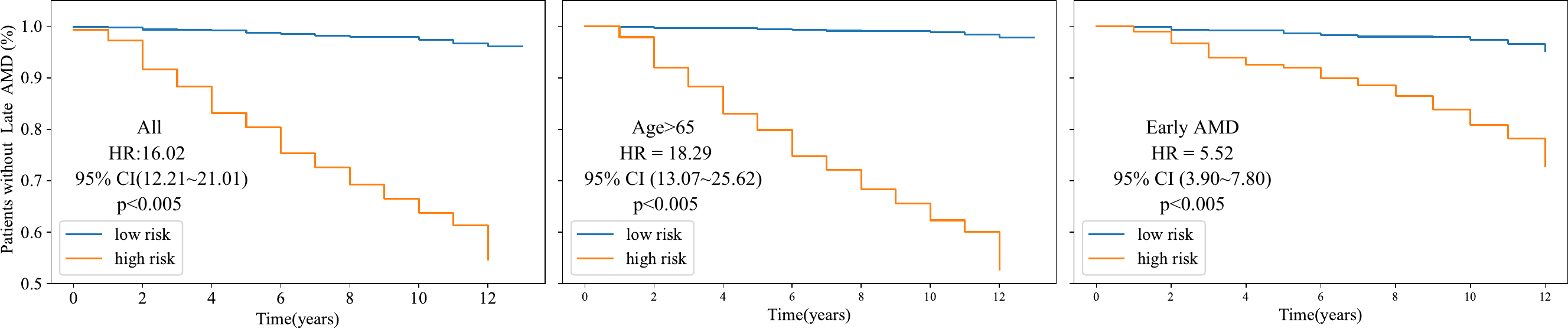}
    \caption{Kaplan–Meier (KM) analysis of AMD survival predictions based on the proposed biomarker in all cases and in subgroups with additional factors. High-risk cases identified by the proposed method may benefit from more intensive interventions at earlier disease stages or in specific patient groups.}
    \label{fig:KM}
\end{figure}

\textbf{Developing a New Biomarker.} We introduce a new biomarker derived from the model's predicted risk, categorizing all cases into two subgroups (\textbf{low-risk} vs \textbf{high-risk}). We use univariate and multivariate Cox proportional-hazards models to evaluate our proposed biomarker alongside other clinical variables, including previously mentioned genetic variants, socio-demographic, and AMD severity score, as well as 10 AMD phenotypes annotated by expert human graders \cite{age2001age}. As shown in Table \ref{tab:multicox}, after selecting significant factors ($p<0.05$) in univariate analysis, our proposed biomarker remains the strongest biomarker among other variables in the multivariate analyses. This finding highlights the effectiveness of the new biomarker. Furthermore, as illustrated in Fig.~\ref{fig:KM}, the proposed biomarker can be combined with other commonly used clinical variables to better identify high-risk patients at early AMD stages or other subgroups (such as old subgroup), thereby offering greater potential for targeted interventions and improved patient outcomes.
%
%

\section{Conclusion}
In conclusion, our proposed AMD-Mamba framework integrates color fundus images, genetic variants, and socio-demographic variables. This approach not only demonstrates robust predictive performance but also introduces a novel biomarker with independent prognostic value, thereby facilitating timely interventions for high-risk individuals. In clinical practice, these findings hold significant promise for improving patient outcomes and guiding more personalized management of AMD.

\vspace{1em}

\noindent\textbf{Acknowledgment.} This work was supported by the National Eye Institute of the National Institutes of Health (NIH) under grant numbers R21EY035296; National Science Foundation under grant numbers 2145640. This research was supported by the NIH Intramural Research Program, National Library of Medicine, and the National Eye Institute. The content is solely the responsibility of the authors and does not necessarily represent the official views of the NIH.

%
%
%
%
\bibliographystyle{unsrtnat}
\bibliography{preprint}






\newpage
\appendix
\setcounter{table}{0}
\setcounter{figure}{0}
\renewcommand\figurename{Supplementary Figure} 
\renewcommand\tablename{Supplementary Table}



\end{document}